\documentclass[11pt]{article}
\title{Sheep Collisions: the Good, the Bad, and the TBI}
\author{
Michael Courtney\thanks{ 
Ballistics Testing Group, P.O. Box 24, West Point, NY 10996
Michael\_Courtney@alum.mit.edu (The Ballistics Testing Group is not affiliated with the United States Military Academy.)}\/ and
Amy Courtney\thanks{
Department of Physics, United States Military Academy, West Point, NY 10996 Amy.Courtney@usma.edu
}
}

\begin{document}     
\maketitle

\begin{abstract}
The title page of Chapter 9 in {\bf Fundamentals of Physics} (Halliday, Resnick, and Walker, $8^{th}$ Edition, p. 201) shows a dramatic photograph of
two Big Horn sheep butting heads and promises to explain how 
sheep survive such violent clashes without serious injury.  However, 
the answer presented in sample problem 9-4 (p. 213) errs in presuming an interaction time of $0.27 s$ which results in an unrealistically long stopping distance of $0.62 m$.  Furthermore, the assertion that the horns provide necessary cushioning of the blow is inconsistent with the absence of concussions in domestic breeds of hornless sheep.  Results from traumatic brain injury (TBI) research allow acceleration tolerance of sheep to be estimated as $450 g$ facilitating an analysis of sheep collisions that is more consistent with available observations (stopping distance less than $1 cm$, impact time of roughly $2 ms$).  
\end{abstract}

\section{ERRANT EXAMPLE IN POPULAR BOOK}
Sample problem 9-4 in a popular textbook stands out because the title page of Chapter 9 shows a dramatic photograph of
two Big Horn sheep butting heads and asserts that the chapter will describe how 
the sheep can survive such violent clashes without falling to the ground with concussions.\cite{walker1}  Jearl Walker also uses this photo and example in speaking engagements entitled  {\em The Flying Circus of Physics}.\cite{walker2} However, the answer presented in Sample Problem 9-4 is incorrect.  The physics 
seems sound, but the flaw in presuming an interaction time of $t_f=.27 s$ becomes clear when the resulting stopping distance is computed.

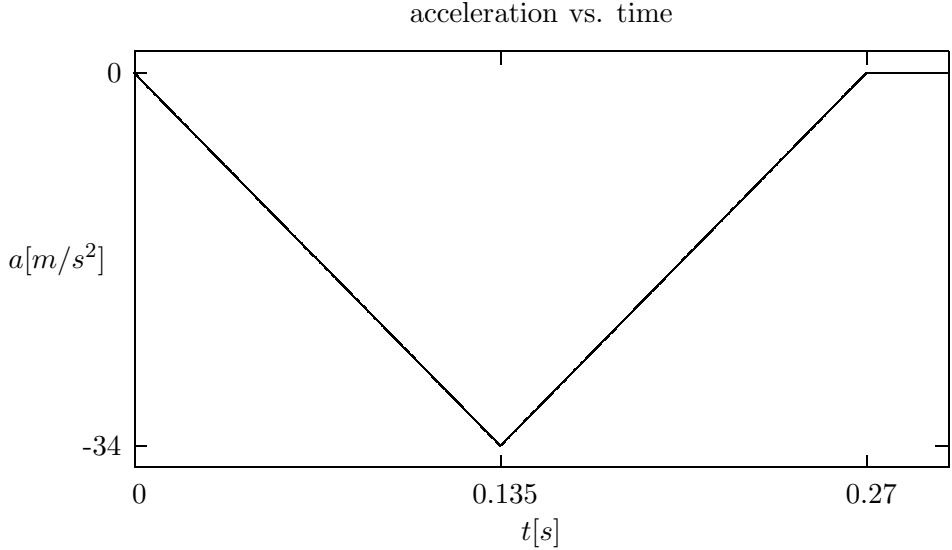
\begin{figure}[hbt]
\label{fig:acceleration}
% GNUPLOT: LaTeX picture
\setlength{\unitlength}{0.240900pt}
\ifx\plotpoint\undefined\newsavebox{\plotpoint}\fi
\begin{picture}(1500,900)(0,0)
\sbox{\plotpoint}{\rule[-0.200pt]{0.400pt}{0.400pt}}%
\put(161.0,157.0){\rule[-0.200pt]{4.818pt}{0.400pt}}
\put(141,157){\makebox(0,0)[r]{-34}}
\put(1419.0,157.0){\rule[-0.200pt]{4.818pt}{0.400pt}}
\put(161.0,743.0){\rule[-0.200pt]{4.818pt}{0.400pt}}
\put(141,743){\makebox(0,0)[r]{ 0}}
\put(1419.0,743.0){\rule[-0.200pt]{4.818pt}{0.400pt}}
\put(161.0,123.0){\rule[-0.200pt]{0.400pt}{4.818pt}}
\put(161,82){\makebox(0,0){ 0}}
\put(161.0,757.0){\rule[-0.200pt]{0.400pt}{4.818pt}}
\put(736.0,123.0){\rule[-0.200pt]{0.400pt}{4.818pt}}
\put(736,82){\makebox(0,0){ 0.135}}
\put(736.0,757.0){\rule[-0.200pt]{0.400pt}{4.818pt}}
\put(1311.0,123.0){\rule[-0.200pt]{0.400pt}{4.818pt}}
\put(1311,82){\makebox(0,0){ 0.27}}
\put(1311.0,757.0){\rule[-0.200pt]{0.400pt}{4.818pt}}
\put(161.0,123.0){\rule[-0.200pt]{307.870pt}{0.400pt}}
\put(1439.0,123.0){\rule[-0.200pt]{0.400pt}{157.549pt}}
\put(161.0,777.0){\rule[-0.200pt]{307.870pt}{0.400pt}}
\put(161.0,123.0){\rule[-0.200pt]{0.400pt}{157.549pt}}
\put(40,450){\makebox(0,0){$a [m/s^2]$}}
\put(800,21){\makebox(0,0){$t [s]$}}
\put(800,839){\makebox(0,0){acceleration vs. time}}
\put(161,743){\usebox{\plotpoint}}
\multiput(161.58,740.89)(0.500,-0.510){1147}{\rule{0.120pt}{0.508pt}}
\multiput(160.17,741.95)(575.000,-584.946){2}{\rule{0.400pt}{0.254pt}}
\multiput(736.58,157.00)(0.500,0.510){1147}{\rule{0.120pt}{0.508pt}}
\multiput(735.17,157.00)(575.000,584.946){2}{\rule{0.400pt}{0.254pt}}
\put(1311.0,743.0){\rule[-0.200pt]{30.835pt}{0.400pt}}
\put(161.0,123.0){\rule[-0.200pt]{307.870pt}{0.400pt}}
\put(1439.0,123.0){\rule[-0.200pt]{0.400pt}{157.549pt}}
\put(161.0,777.0){\rule[-0.200pt]{307.870pt}{0.400pt}}
\put(161.0,123.0){\rule[-0.200pt]{0.400pt}{157.549pt}}
\end{picture}
\caption{\em Acceleration of Big Horn sheep given in sample problem.}
\end{figure}

Fig. 1 %\ref{fig:acceleration}  
shows the acceleration given in the sample problem.  Time $t=0 s$ 
corresponds to the beginning of the collision.  Letting $a_{max} = 34 m/s^2$,
the acceleration decreases linearly from $t=0 s$ until $t_1=\frac{t_f}{2}=0.135 s$
as
\begin{equation}
a_1(t) = -\frac{2a_{max}}{t_f} t. 
\end{equation}
Then the acceleration increases linearly from $t_1=0.135 s$ to $t_f=0.27 s$ as
\begin{equation}
a_2(t) = -2a_{max} + \frac{2a_{max}}{t_f} t. 
\end{equation}

The sheep each have a mass of $90 kg$ so that F(t) is the mass times the acceleration function.
The impulse is calculated to be the area under the force curve or $-\frac{1}{2}m a_{max}t_f = -413 N\cdot s$.  
Since this is the impulse that brings one sheep to a stop, it is equal and opposite to the initial momentum of the sheep, 
$p_i = \frac{1}{2}m a_{max}t_f= 413 kg\cdot m/s$, which implies an initial velocity, 
$v_i = p_i/m = \frac{1}{2}a_{max}t_f = 4.589 m/s$.  (One can also determine
$v_i$ from the average acceleration $a_{ave}= -17 m/s^2$ and the time interval $0.27 s$.)

The explanation in the ``Comment:'' portion of the sample problem asserts that the 
horns of the sheep cushion the impact by increasing the duration of the collision
to $t_f = 0.27 s$.  Supposedly, injury is prevented, because the long duration reduces the forces to a level that the sheep can tolerate.
However, for an impact duration of $t_f = 0.27 s$, the horns would have to extend forward beyond the skull for a distance at least as large as the 
stopping distance, which we will show is unrealistic.

\subsection{Velocity from integrating the acceleration}

The velocity decreases from $t=0 s$ until $t_1=0.135 s$ as
\begin{eqnarray}
v_1(t) = v_i + \int_{0}^t a_1(t) dt 
= v_i + \int_{0}^t \frac{-2a_{max}}{t_f} t dt \nonumber \\
= \frac{1}{2}a_{max}t_f - \frac{a_{max}}{t_f} t^2. 
\end{eqnarray}
Evaluating at $t = t_1 = \frac{t_f}{2} = 0.135 s$ gives $v_{2i} = \frac{1}{4} a_{max}t_f = 2.295 m/s$.

The velocity then decreases from $t_1=0.135 s$ to $t_f=0.27 s$ as
\begin{eqnarray}
v_2(t) = v_{2i} + \int_{t_1}^t a_2(t) dt 
= v_{2i} + \int_{t_1}^t \left(-2a_{max}+ \frac{2a_{max}}{t_f} t\right) dt \nonumber \\
v_2(t) = v_{2i}  
+ 2a_{max} \frac{t_f}{2} - \frac{a_{max}}{t_f}\left(\frac{t_f}{2}\right)^2 - 2a_{max} t + \frac{a_{max}}{t_f} t^2. 
\end{eqnarray}
which substituting in $v_{2i} = \frac{1}{4} a_{max}t_f = 2.295 m/s$ simplifies to
\begin{eqnarray}
v_2(t) = a_{max} t_f - 2 a_{max} t + \frac{a_{max}}{t_f} t^2 . 
\end{eqnarray}
As expected, $v_2(t) = 2.295 m/s$ at $t = \frac{t_f}{2} = 0.135 s$, and $v_2(t) = 0 m/s$ at $t = t_f = 0.27 s$.

\begin{figure}[hbt]
%\vspace{3.0in}
\label{fig:velocity}
\input{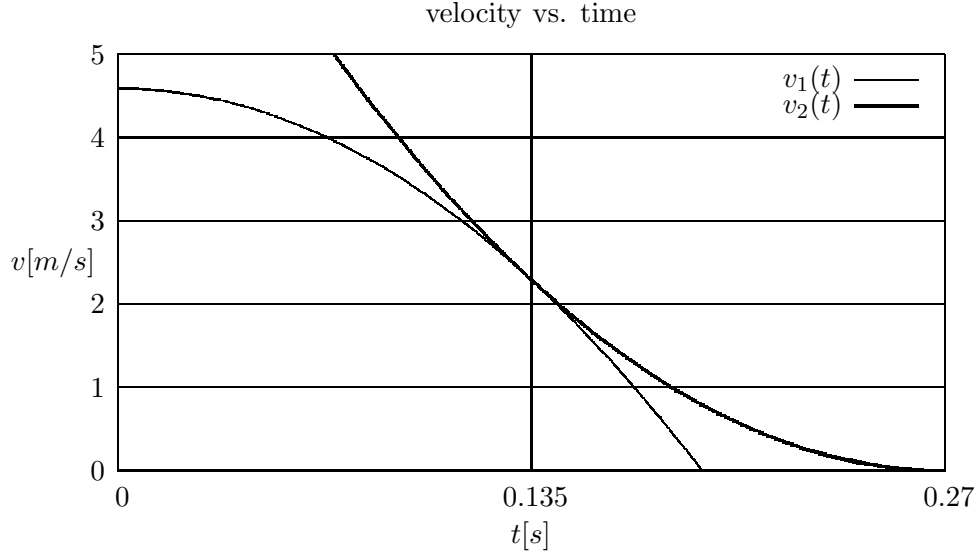}
\caption{\em Velocity of sheep during collision given model acceleration.  Note that the velocity 
is $v_1(t)$ for $0 \leq t \leq 0.135 s$ and $v_2(t)$ for $0.135 s \leq t \leq 0.27 s$.
}
\end{figure}
Fig. 2 %\ref{fig:velocity} 
shows the velocity of one sheep.  As expected, 
the velocity decreases monotonically from $t=0 s$ until $t_f=0.27 s$.  However, 
unlike many introductory physics problems (those with constant acceleration), 
the velocity is not a linear function of time.

\subsection{Position from integrating the velocity}

The displacement during impact increases from $t=0 s$ to $t_1= \frac{t_f}{2} = 0.135 s$ as 
\begin{eqnarray}
x_1(t) = v_i t + \int_{0}^t \frac{-a_{max}}{t_f} t^2 dt
= \frac{1}{2}a_{max}t_f t - \frac{a_{max}}{3t_f} t^3. 
\end{eqnarray}
Evaluating this at $t = t_1 = \frac{t_f}{2} = 0.135 s$ gives $x_{2i} = \frac{5}{24} a_{max}t_f^2 = 0.5162 m$.

The position continues to  increase from $t_1=\frac{t_f}{2} = 0.135 s$ 
to $t_f=0.27 s$ as 
\begin{eqnarray}
x_2(t) = x_{2i} + \int_{t_1}^t \left(a_{max}t_f - 2a_{max} t + \frac{a_{max}}{t_f} t^2\right) dt \nonumber \\
x_2(t) = -\frac{1}{12}a_{max}t_f^2 +  a_{max}t_f t - a_{max}t^2 + \frac{a_{max}}{3t_f} t^3.
\end{eqnarray}
$x_1(t)$ and $x_2(t)$ are shown in Fig. 3.
Evaluating $x_2(t)$ at $t = t_f$ gives the total stopping distance:
\begin{eqnarray}
x_2(t_f) = \frac{1}{4}a_{max} t_f^2 = \frac{1}{2}v_i t_f = 0.6195 m.
\end{eqnarray}
Fig. 3 %\ref{fig:position} 
shows the displacement after impact.

\begin{figure}[hbt]
%\vspace{3.0in}
\label{fig:position}
\input{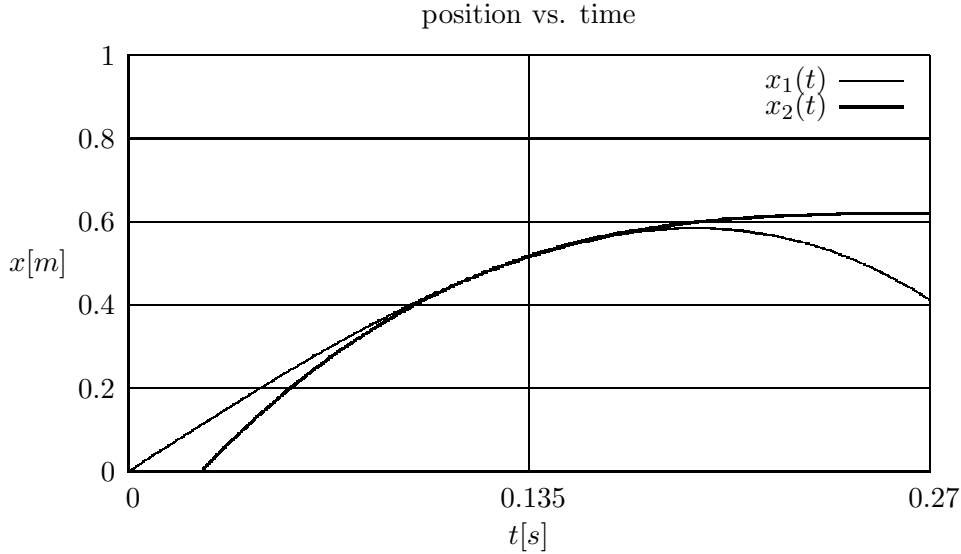}
\caption{\em Position of sheep during collision given model acceleration.  Note that the position 
is $x_1(t)$ for $0 \leq t \leq 0.135 s$ and $x_2(t)$ for $0.135 s \leq t \leq 0.27 s$.
}
\end{figure}

\subsection{Discussion of concussion and brain injury}
The assertion that horns
cushion the blow by spreading out the force over a time interval of $0.27 s$
results in an unrealistic stopping distance of $0.6195 m$, because the horns
do not extend forward or flex sufficiently 
to provide this cushion.  The stopping distance 
provided by the horns is likely much smaller resulting in much larger forces and accelerations.

It is well recognized in the traumatic brain injury (TBI) 
literature\cite{shaw} 
that certain animals(such as woodpeckers and head-butting ruminants) are much more resistant than humans to TBI and that this
resistance to TBI is not simply explained by introductory mechanics.\cite{gibson}
The detailed biomechanical reasons for this resistance to TBI are not well-understood, but there is
some evidence to suggest that ruminants that engage in pre-mating head butting rituals are not only more resistant to TBI from externally imposed 
forces, accelerations, and blunt-force trauma. They are also more resistant to TBI induced 
from an internal ballistic pressure 
wave originating in the thoracic cavity.\cite{courtney} 
This suggests that resistance to TBI in these species has a neurological 
as well as a biomechanical basis.

\subsection{A simpler approach}
Few textbook readers are likely to perform the above integrations necessary to 
determine the stopping distance.  However, a reasonable estimate of the stopping distance 
is possible simply using the average acceleration and assuming (wrongly, of course) that
the accleration is constant and estimating the stopping distance from kinematics (or the 
Work-Energy theorem).

Since $v_i = 4.589 m/s$ and $v_f = 0 m/s$, assuming a constant acceleration gives 
$v_{ave}= v_i/2 = 2.295 m/s$.  The estimated stopping distance is $x = v_{ave} t_f = 0.6195 m$.
Why is the stopping distance estimated from an erroneous assumption the same as the rigorously 
computed stopping distance?  The symmetry of the acceleration about $t = 0.135 s$ results in a symmetry 
in $v(t)$ so that $v_{ave}= v_i/2 = 2.295 m/s$ even though the acceleration is not constant.  
This is a happy accident that could not be
inferred with confidence without well-developed intuition.\cite{symmetry}
However, estimates of stopping distances 
from assuming a constant acceleration are often within a factor of two of rigorous calculations.

\section{A MORE REASONABLE ANALYSIS}
Having raised domestic sheep, we have observed that a number of hornless sheep breeds  
engage in pre-mating dominance rituals of head-butting without falling to the ground with concussions.\cite{krouse}
This is also observed by many shepherds.
If breaking off a horn resulted in great danger to Big Horn sheep because horns are critical in spreading
out the momentum change so that the forces and accelerations are smaller,\cite[Sample Problem 9-4]{walker1} then a higher incidence of concussion would be seen in the hornless sheep breeds.

Head butting ruminants are more resistant to concussion than humans, but firm quantitative thresholds for tramautic brain injury (TBI) in humans remain a matter of debate and ongoing research.  Both motorcycle and combat helmet standards can be expressed in terms of allowable model head accelerations for a specified impact velocity under certain test conditions.  (A moving model head wearing the helmet impacts a stationary object and the head acceleration is measured.)  

For example, the US Army's standard for their advanced combat helmet is $150 g$ when impacting a hard surface at 10 ft/s.\cite{ach}  The Department of Transportation (DOT) standard for motorcycle helmets is roughly $250 g$ for high-energy impacts and most DOT approved helmets produce under $200 g$ for commonly encountered impacts.\cite{dot} An oft-cited car crash study estimated the acceleration tolerance in humans as $220 g$ for impacts with a duration of $2 ms$\cite{ono}.  While it is not well established that these acceleration thresholds accurately predict the presence or absence of mild TBI, these standards represent an accepted range of acceleration tolerance in humans.\cite{zhang}  

Using the approximation of constant acceleration, combining an acceleration threshold with an impact velocity allows a computation of a minimum stopping distance from kinematics:  
\begin{eqnarray}
d = \frac{v_i^2}{2 a},
\end{eqnarray}
where $d$ is the stopping distance, $v_i$ is the impact velocity, and $a$ is the acceleration threshold.
For an acceleration of $150 g$ and an impact velocity of $10 ft/s$ ($2.5 m/s$), the minimum stopping distance is $0.01 ft$ ($0.25 cm$).  

The impact velocity of Big Horn sheep might be as large as $v_i = 8.9 m/s$.\cite{NG} The acceleration tolerance in sheep can be estimated as $450 g$, which  is consistent with the Gibson scaling rule\cite{gibson}.  This approach yields a minimum stopping distance of roughly $0.009 m$.  This is much smaller than the sample problem's stopping distance of $0.62 m$, and can be accommodated even in hornless sheep.  This suggests the greater resistance of sheep to TBI (higher acceleration threshold) rather than horns is the dominant factor in sheep survival of head butting collisions.\cite{crude}  This analysis implies an interaction time of roughly $0.002 s$, much shorter than the sample problem interaction time of $0.27 s$.

\section{CONCLUSION}
In conclusion, both the theoretical analysis (computation of stopping distance) and the experimental evidence (observations
in hornless domestic sheep) contradict the assertion that the sheep's resistance 
to concussion is related to the horn spreading out the impulse over a time interval as long as $0.27 s$.  Stopping distances shorter than $0.01 m$ are consistent with reasonable estimates of acceleration threshold and impact velocity.  There may be biomechanical properties within the sheep heads that provide for a cushion of roughly $1 cm$, but their resistence to concussions also seems to have a neurological basis.  The question of sheep resistance to head butting injury makes for an interesting and illuminating example in introductory Physics, but realistic values of the interaction time and distance must be used to compute reasonable values of the forces and accelerations involved.


\begin{thebibliography}{99}

\bibitem{walker1} David Halliday, Robert Resnick, Jearl Walker, {\it Fundamentals of Physics} (Wiley, New York, 2007), $8^{th}$ ed. p. 201, p. 213.

\bibitem{walker2} There is also a qualitatitative description in the book by the same name.  Jearl Walker, {\it The Flying Circus of Physics}, (Wiley, New York, 2006), $2^{nd}$ ed.

\bibitem{shaw} Nigel A. Shaw, ``The Neurophysiology of Concussion,'' Progress in Neurobiology {\bf 67}, 281-344 (2002).

\bibitem{gibson} L.J. Gibson, ``Woodpecker pecking: How woodpeckers avoid brain injury,'' Journal of Zoology {\bf 270}, 462-465 (2006). Small brain size, short impact duration, and brain orientation within the skull combine to contribute to woodpecker acceleration tolerance.

\bibitem{courtney} See Figure 6, Michael Courtney, Amy Courtney, 
``Ballistic pressure wave contributions to rapid incapacitation in the Strasbourg goat tests,''  
arxiv.org/ftp/physics/papers/0701/0701267.pdf 

\bibitem{symmetry} It is a general result that 
$v_{ave}= v_i/2$ for stopping problems where the acceleration is symmetric about the midpoint in time.  

\bibitem{krouse}Krouse P, ``Young Growers Go After Niche of Selling Right to Consumers,'' {\it Cleveland Plain Dealer}, October 23, 2002, p. C1.  The sheep pictured are two hornless rams that regularly engaged in head-butting.  Also see: {\it Better Physics Through Farming} at www.ballisticstestinggroup.org/mwc.htm

\bibitem{ach} B.J. McEntire, P. Whitley, ``Blunt Impact Performance Characteristics of the
Advanced Combat Helmet and the Paratrooper and Infantry Personnel Armor System
for Ground Troops Helmet,'' U.S. Army Aeromedical Research
Laboratory. USAARL Report No. 2005-12 (2005).

\bibitem{dot} The DOT standard for motorcycle helmets can be translated into an acceleration for a specific impact velocity.  See 
www.motorcyclistonline.com/gearbox/motorcycle\_helmet\_review/ 
for a discussion of motorcycle helmet testing and standards.

\bibitem{ono} K. Ono, A. Kikuchi, M. Nakamura, H. Kobayashi, and N. Nakamura, ``Human head tolerance to sagittal
impact: reliable estimation deduced from experimental
head injury using subhuman primates and human cadaver
skulls,'' In {\it Proceedings of the 24th Stapp car crash conference}, 101–160. Warrendale, PA: Society of Automotive
Engineers (1980). 

\bibitem{zhang}Studies of helmet to helmet collisions in professional football suggest a threshold range of $66g - 106g$ for impacts of roughly $10 ms$ duration.  Liying Zhang, King H. Yang, and Albert I. King, ``A Proposed Injury Threshold for Mild Traumatic Brain Injury,'' {\it Journal of Biomechanical Engineering} {\bf 126}, 226-236 (2004).  

\bibitem{NG}{See animals.nationalgeographic.com/animals/mammals/rocky-mountain-bighorn-sheep.html}

\bibitem{crude}{These estimates are rough, but  at least represent collision parameters consistent with each other and available observations.
}


\end{thebibliography}
\end{document}